\documentclass[preprint,12pt]{elsarticle}
\usepackage{amssymb}

\journal{Nuclear Physics B}

\begin{document}

\begin{frontmatter}

\title{Investigating bounds on decoherence in quantum mechanics via 
B and D-mixing}

\author[a]{Alexander Lenz,}
\author[a]{David Hodges,}
\author[a]{Daniel Hulme,} 
\author[a]{Sandra Kvedaraite,}
\author[a]{Jack Richings,}
\author[a]{Jian Shen Woo}
\author[a]{and Philip Waite}

\address[a]{Institute for Particle Physics Phenomenology, Durham University, 
\\ 
DH1 3LE Durham, United Kingdom
\\
alexander.lenz@durham.ac.uk}

\begin{abstract}
We investigate bounds on decoherence in quantum mechanics by studying $B$ and 
$D$-mixing observables, making use of many precise new measurements, 
particularly from the LHC and B factories. In that respect we show that
 the stringent bounds obtained by a different group in 2013 rely on 
unjustified assumptions. Finally, we point out which
experimental measurements could improve the decoherence bounds considerably.
\end{abstract}

\begin{keyword}
Quantum Mechanics, Decoherence, Mixing, B-Physics, CP violation
\end{keyword}

\end{frontmatter}

%--------+---------+---------+---------+---------+---------+---------+---------+
%
%  Section: Introduction
%
%--------+---------+---------+---------+---------+---------+---------+---------+
\section{Introduction}
\label{sec:intro}
The interpretation of quantum mechanics is still an open problem, see for 
example
the very recent discussion of the quantum pigeonhole in \cite{pigeon}.
Many of the related questions go back to the effect of entanglement that was
studied in 1935 by Einstein, Podolsky and Rosen
\cite{Einstein:1935rr}. Numerous different systems have been investigated in 
that respect.
It is also interesting to test the validity of the foundations of quantum
mechanics in systems that are usually used to search for physics beyond the 
standard model.
Therefore we  discuss here the mixing of neutral mesons, which is a well-known 
and well-studied quantum mechanical effect, see e.g. \cite{Datta:1986ut}
for an early discussion of $B$-mesons in that respect. 
It leads to the fact that the neutral mesons that are propagating 
in space-time are described by the mass eigenstates, e.g. $B_H$ and $B_L$, which 
are linear combinations of the flavour eigenstates defined by the quark 
content, e.g. $B_d = (\bar{b}d)$ and  $\bar{B}_d = (b \bar{d})$:
\begin{eqnarray}
B_H & = & p B_d - q \bar{B}_d \; ,
\\
B_L & = & p B_d + q \bar{B}_d \; .
\end{eqnarray}
In these systems we have the following observables: the mass difference
$\Delta M = M(B_H) - M(B_L)$, the decay rate difference $\Delta \Gamma
= \Gamma (B_L) - \Gamma (B_H)$ and semi-leptonic CP asymmetries, which can be expressed
as $a_{sl} = 2 ( 1- |q/p|)$.
\\
For the two neutral $B$-meson systems, we now have quite precise
experimental numbers for the mixing observables. The HFAG 2014 \cite{HFAG}
%\footnote{Based on the measurements
%of $\Delta M_d$ by xxxx\cite{},
%of $\Delta M_s$ by xxxx\cite{},
%of $\Delta \Gamma_d$ by xxxx\cite{} 92,
%of $\Delta \Gamma_s$ by xxxx\cite{} 93,
%of $a_{sl}^d$ by xxxx\cite{}
%and
%of $a_{sl}^s$ by xxxx\cite{}.} 
values read:
\begin{equation}
\begin{array}{|c||c|c|}
\hline
              & B_s   &B_d   
\\
\hline
\hline
\Delta M   &(17.761  \pm 0.022) \, \mbox{ps}^{-1}& (0.510\pm 0.003)\, \mbox{ps}^{-1}
\\
\hline
\Delta \Gamma  &(0.091  \pm 0.008) \, \mbox{ps}^{-1} &(0.001  \pm 0.010) \, \Gamma 
\\
\hline
\Gamma  &\frac{1}{1.512 \pm 0.007} \, \mbox{ps}^{-1}&\frac{1}{1.519 \pm 0.005} \, \mbox{ps}^{-1}
\\
\hline
a_{sl}  &-0.0077 \pm 0.0042 &-0.0009 \pm 0.0021 
\\
\hline
\end{array} \; ,
\end{equation}
where $\Gamma$ denotes the total decay rate of the neutral $B$-mesons.
The experimental numbers for $\Delta M_d$, $\Delta M_s$ and 
$\Delta \Gamma_s$ are now dominated by LHC measurements. 
The most precise values were obtained 
for $\Delta M_d$ by LHCb \cite{Aaij:2012nt},
for $\Delta M_s$ by LHCb \cite{Aaij:2013mpa}
and
for $\Delta \Gamma_s$ by ATLAS \cite{Aad:2014cqa},
CMS \cite{CMS:2014jxa}
and 
LHCb \cite{Aaij:2013oba}.
Here no entanglement effects are expected to occur, hence we use these values 
as independent inputs for our investigations of decoherence.
For the semi-leptonic asymmetries and $\Delta \Gamma_d$ we do not yet have 
clear experimental evidence for a non-zero value; only some bounds 
are available.
Thus we give for completeness the corresponding
standard model predictions 
\cite{Lenz:2011ti,Lenz:2006hd,Ciuchini:2003ww,Beneke:2003az,Beneke:1998sy}
of these quantities:
\begin{equation}
\begin{array}{|l||l|l|}
\hline
              & B_s   &B_d   
\\
\hline
\hline
\Delta \Gamma  &(0.087  \pm 0.021)\, \mbox{ps}^{-1}&(0.0029 \pm 0.0007) \, \mbox{ps}^{-1}
\\
\hline
a_{sl}  &(1.9 \pm 0.3) \cdot 10^{-5}&(-4.1 \pm 0.6) \cdot 10^{-4}
\\
\hline
\end{array}  \; .
\end{equation}
In the neutral $D$-meson system, typically the mixing parameters $x$, $y$ and 
$|p/q|$ are determined
directly \cite{HFAG}:
\begin{eqnarray}
x := \frac{\Delta M}{\Gamma} & = & 0.41^{+0.14}_{-0.15} \; ,
\\
y := \frac{\Delta \Gamma}{2\Gamma} & = & 0.63^{+0.07}_{-0.08} \; ,
\\
\left| \frac{p}{q} \right| & = & 0.93^{+0.09}_{-0.08} \; .
\end{eqnarray}

\section{Decoherence in B-mixing}
\label{sec:deco}
At the B-factories B-mesons were typically produced via the decay of the 
$\Upsilon(4s)$ resonance, thus also producing entangled $\bar{B}_d B_d$ pairs.
To describe decoherence in $B_d$-mixing, semileptonic decays of the neutral 
B-mesons were investigated, e.g. $\bar{B}_d \to l^- \bar{\nu}_l X$ and
${B}_d \to l^+ {\nu}_l X$. 
If no mixing occurs, one gets from semi-leptonic decays events with one 
positively charged and one negatively charged lepton - so-called 
opposite-sign leptons. If mixing is taken into account one can also get
events with two positively or two negatively charged leptons, so-called
same-sign leptons.
Following \cite{Bertlmann:1996at} we define the ratio of like-sign dilepton decays of a 
neutral $B$-meson to opposite-sign dilepton events, $R$,
(based on the investigations in \cite{Carter:1980tk,Bigi:1981qs}), as
\begin{eqnarray}
R & = & \frac{N^{++} + N^{--}}{N^{+-} + N^{-+}} \; .
\end{eqnarray}
$N^{++}$ denotes the events with two positively charged leptons in the final states and
so on.
In \cite{Bertlmann:1996at} Bertlmann and Grimus used the parameter $\zeta$ to
describe decoherence effects in quantum mechanics  
in a phenomenological manner. $\zeta = 0$ corresponds 
to the familiar case of quantum mechanics, while $\zeta = 1 $ describes a case
where no quantum mechanical interference effects are occurring at all - corresponding
to Furry's hypothesis \cite{Furry}.
The general expression for $R$ in terms of the mixing parameters $x$, $y$ and 
$|p/q|$ then reads \cite{Bertlmann:1996at}
\begin{eqnarray}
R &=&
\frac12 \left( 
\left| \frac{p}{q} \right|^2
+
\left| \frac{q}{p} \right|^2
\right)
\frac{x^2+y^2       +\zeta \left[ y^2 \frac{1+x^2}{1-y^2} + x^2 \frac{1-y^2}{1+x^2} \right]}
     {2 + x^2 - y^2 +\zeta \left[ y^2 \frac{1+x^2}{1-y^2} - x^2 \frac{1-y^2}{1+x^2} \right]} \; .
\end{eqnarray}
This formula can be written more in the more compact form
\begin{eqnarray}
R & = & 
R_0 \frac{1}{\sqrt{1-a_{sl}^2}} \frac{1+\alpha \zeta}{1+\beta \zeta} \; ,
\label{Master}
\end{eqnarray}
with
\begin{eqnarray}
R_0 & = &\frac{x^2+y^2}{2 + x^2 - y^2} \; ,
\\
\alpha & = & \frac{y^2(1+x^2)^2 + x^2 (1-y^2)^2}{(x^2+y^2)(1+x^2)(1-y^2)} \; ,
\\
\beta & = & \frac{y^2(1+x^2)^2 - x^2 (1-y^2)^2}{(2+x^2-y^2)(1+x^2)(1-y^2)} \; ,
\\
a_{sl} & = &  \frac{N^{++} - N^{--}}{N^{++} + N^{--}} 
=
\frac{ \left| \frac{p}{q} \right|^2 - \left| \frac{q}{p} \right|^2}
     { \left| \frac{p}{q} \right|^2 + \left| \frac{q}{p} \right|^2} \; .
\end{eqnarray}
$R_0$ can also be expressed in terms of the mixing probability
\begin{equation}
\chi =  \frac{x^2 + y^2}{2(1+x^2)} \; ,
\label{mixprob}
\end{equation}
as
\begin{equation}
R_0 = \frac{\chi}{1-\chi} \, .
\end{equation}
The current values of $\chi$ from PDG \cite{PDG} or HFAG \cite{HFAG} are not measured 
directly but derived via  Eq.(\ref{mixprob}) from the direct measurements of $\Delta M$, 
$\Delta \Gamma$ and $\Gamma$.
Historically $R$ was approximated by $R_0$ and used to extract the mixing
probability, see e.g. \cite{Albrecht:1993gr,Bartelt:1993cf}.
\\
Eq.(\ref{Master}) represents our master equation. The numerical values of 
the coefficients in this equation can be calculated quite precisely  by using
the most recent experimental numbers from HFAG \cite{HFAG}. For the 
heavy neutral mesons we get:
\begin{equation}
\label{parameter1}
\begin{array}{|c||c|c|c|}
\hline
              & B_s  = \left(\bar{b}s\right)  &B_d = \left(\bar{b}d\right)  & 
D^0 = \left(c \bar{u}\right)  
\\
\hline
\hline
R_0    & 0.997247 \left(1 \pm 2.7 \cdot 10^{-5}\right)& 0.2308 (1 \pm 0.010)& 0.319 \left(1^{+0.27}_{-0.25} \right)
\\
\hline
\alpha & 6.14^{+0.88}_{ -0.80} \cdot 10^{-3} &  0.6249^{+0.0032}_{-0.0032} & 1.51^{+0.31}_{-0.24}
\\
\hline
\beta  & 3.37^{+ 0.88}_{- 0.80} \cdot 10^{-3} &-0.14424^{+0.00077}_{-0.00076}& 0.39^{+0.24}_{-0.17}
\\
\hline
\frac{1}{\sqrt{1-a_{sl}^2}} & 
1+ 3.0^{+ 4.1}_{-2.4} \cdot 10^{-5}&
1+ 4.1^{+41.0}_{-4.1} \cdot 10^{-7}&
 1.011^{+0.043}_{-0.010}
\\
\hline
\end{array} \; .
\end{equation}
The above coefficients are known quite precisely for the neutral $B$-system, while there are still
sizable uncertainties in the $D$-meson systems.
If quantum mechanics holds, i.e. $\zeta = 0$,  we predict (by using the measured values of $\Delta \Gamma$, 
$\Delta M$, $\Gamma$ and $a_{sl}$) the following values for the ratio $R$:
\begin{equation}
\label{RQM}
\begin{array}{|c||c|c|c|c|}
\hline
              & B_s   &B_d   & D^0   
\\
\hline
\hline
R^{\rm QM}      & 0.997277^{+0.000049}_{-0.000036} & 0.2308 \pm  0.0024 &  0.322^{+ 0.089}_{- 0.079}
\\
\hline
\end{array} \; .
\end{equation}
These numbers can be compared to the measured values of $R$ stemming from ARGUS (1993) 
\cite{Albrecht:1993gr} 
and CLEO (1993) \cite{Bartelt:1993cf} for the $B_d$-system. To our knowledge there are no measurements
available for the $B_s$ or the $D$-meson system:
\begin{equation}
\label{REXP}
\begin{array}{|c||c|c|c|c|}
\hline
              & B_s   &B_d   & D^0   
\\
\hline
\hline
R      & -       & 0.194 \left( 1 \pm 0.424 \right) %\cite{Albrecht:1993gr} 
& -
\\
       &         &  0.187 \left( 1^{+0.278}_{-0.239} \right) %\cite{Bartelt:1993cf} 
&
\\
\hline
\end{array}
\; .
\end{equation}
The first entry in the table is from ARGUS \cite{Albrecht:1993gr}, while the 
second is from CLEO (1993) \cite{Bartelt:1993cf}.
A significant deviation of $R^{\rm QM}$ from the measured value $R$ would point 
towards a violation of quantum mechanics, that could be described by a non-vanishing value
of $\zeta$. This can be expressed as
\begin{equation}
\zeta = \frac{\frac{R}{R_0} \sqrt{1-a_{sl}^2} - 1}
        {\alpha - \beta \frac{R}{R_0} \sqrt{1-a_{sl}^2}} \; .
\label{zeta}
\end{equation}
If quantum mechanics holds, then Eq.(\ref{Master}) states 
$R = R_0/\sqrt{1-a_{sl}^2}$ and thus  Eq.(\ref{zeta}) gives
 $\zeta =0$, as expected.
Using the experimental values for $R$ from
\cite{Albrecht:1993gr} or \cite{Bartelt:1993cf} we get from 
Eq.(\ref{zeta}) the following bounds on $\zeta$:
\begin{eqnarray}
\zeta & = &  - 0.26^{+0.30}_{-0.28} 
\; , 
\label{bound_zeta}
\\
\zeta & = &  -0.21^{+0.46}_{-0.53} 
\; .
\label{bound_zeta2}
\end{eqnarray}
Eq.(\ref{bound_zeta}) has been calculated using the value of $R$ measured by
 ARGUS \cite{Albrecht:1993gr}, whereas Eq.(\ref{bound_zeta2}) is derived 
from the CLEO value \cite{Bartelt:1993cf}.
The central value for $\zeta$ is slightly negative, but the value $\zeta = 0 $ is 
within the one standard deviation region of the measured value of $R$, thus no decoherence
effects can be seen yet in the neutral $B_d$-system. 
Total decoherence, i.e. $\zeta = 1$, is thus excluded by about four standard deviations.
The uncertainty of the extracted
value for $\zeta$ is completely dominated by the uncertainty in $R$.
At this point, however,  some caution is necessary. It was shown in \cite{Dass:1997dd,Bertlmann:1997ei}
that bounds on $\zeta$ depend on the basis (flavour or mass basis) used for describing the
neutral B-mesons. Thus quantitative statements describing the deviation
from decoherence have to be taken with some care. 
\\
Using the old experimental inputs from Bertlmann and Grimus \cite{Bertlmann:1996at} we get 
\begin{equation}
\zeta = -0.16^{+0.30}_{-0.31}  \; .
\end{equation}
The shift in the central value stems from the fact that at that time all the mixing parameters 
were known much less precisely. The uncertainty, which is dominated by $R$, stayed more or less the same,
because we use the same value for $R$. 
Unfortunately there are no new measurements of the ratio $R$ in the $B_d$-system 
available and there exists no measurement at all in the $B_s$ or $D$-system.
Thus we are limited by the experimental accuracy stemming from 1993.
Here any new experimental investigation of  $R$ would be very helpful, using for
example
the huge data set collected by the B-factories. 
In that respect it is of course interesting to ask, what experimental precision in
$R$ would result in what bound on $\zeta$? We find:
\begin{equation}
\begin{array}{|c||c|c|c|c|c|c|}
\hline
\delta R & \pm 25 \%  & \pm 10 \%  & \pm 5 \%  & \pm 2 \%  & \pm 1 \% & \pm 0.5 \%
\\
\hline
\hline
\delta \zeta & ^{+1.15}_{-1.07} & ^{+0.45}_{-0.44} & 
               ^{+0.23}_{-0.22} & \pm 0.10        & \pm 0.06 & \pm 0.05  
\\
\hline
\end{array} \; .
\end{equation} 
With a precision of $1 \%$ in $R$, the current uncertainties in $x$, $y$ and $|p/q|$ 
dominate the uncertainty in $\zeta$.
\\
The Belle Collaboration  performed in \cite{Go:2007ww} a time-dependent analysis of semi-leptonic
B-decays and obtained a very strong bound on decoherence:
\begin{equation}
\zeta^{\rm Belle} = 0.029 \pm 0.057 \; .
\end{equation}
It is, however, not completely clear how to relate $\zeta^{\rm Belle}$ to our time-integrated
analysis. Moreover, Belle neglected $\Delta \Gamma_d$ and $a_{sl}$, whose experimental
values can still be in the percentage range, which is comparable to the uncertainty in 
$\zeta^{\rm Belle}$.

\section{The analysis of Alok and Banerjee from 2013}

In \cite{Alok:2013sca} it was tried to avoid the experimental
short-comings related to the almost unknown value of $R$ by a trick: 
because Eq.(\ref{Master}) seems to indicate $R \approx R_0$, it was assumed in  Eq.(\ref{zeta}) 
that $R = R_0$. This is equivalent to starting from
\begin{equation}
R = \frac{\chi}{1-\chi} \; ,
\end{equation}
with $\chi$ defined in Eq.(\ref{mixprob}),
in order to investigate bounds on decoherence.
With that approximation we get
\begin{equation}
\zeta = \frac{ \sqrt{1-a_{sl}^2} - 1}
        {\alpha - \beta  \sqrt{1-a_{sl}^2}}
\label{zeta_wrong}
\; ,
\end{equation}
an equation that does not depend at all on $R$ and has all coefficients
quite precisely known.
Taking Eq.(\ref{zeta_wrong}) as a starting point we obtain  the following
strong bounds on $\zeta$:
\begin{eqnarray} 
\zeta (B_d) &=& -0.53^{+ 0.53}_{- 5.32} \cdot {10^{-6}} \; ,
\\  
\zeta (B_s) &=& -0.0107^{+0.0085}_{- 0.0149} \; .
\end{eqnarray}
Thus Eq.(\ref{zeta_wrong}) leads to very stringent constraints on decoherence in
quantum mechanics, which are orders of magnitude better than the ones obtained in
Eq.(\ref{bound_zeta}). Alok and Banerjee found that total decoherence is excluded 
by 34 standard deviations in the $B_d$-system and by 24 - 31 standard deviations in the 
$B_s$-system. 
\\
However, the first thing to notice when studying Eq.(\ref{zeta_wrong}) is that 
now the value of $\zeta$ depends purely on the experimental value of $a_{sl}$ -
a result which is in contradiction
with the definition of $a_{sl}$, which is independent of $\zeta$.
\\
Secondly, one 
now obtains also a non-vanishing $\zeta$-value if one takes the standard 
model values for $a_{sl}^{s,d}$, which of course assume the validity of quantum mechanics:
\begin{eqnarray}
\zeta \left( a_{sl}^{d, \rm SM} \right) & = & -1.09^{+0.30}_{-0.34} \cdot 10^{-7} \; ,
\\
\zeta \left( a_{sl}^{s, \rm SM} \right) & = & -6.53^{+2.87}_{-4.62} \cdot 10^{-8} \; .
\end{eqnarray}
So we get a violation of quantum mechanics - albeit a tiny one - 
even if we take standard model predictions for the semi-leptonic
asymmetries. This is clearly a contradiction and points to this method 
being invalid.
\\
Thirdly, Alok and Banerjee also found  a bound in the $B_s$-system, even though there was no
measurement at all of $R$ available in this system and $R$ is the only quantity sensitive to
decoherence effects.
\\
Finally, it is clear that one cannot simply equate $R$ and $R_0$ without
taking into account the accuracy of this approximation as well as  
the independent uncertainties of $R$ and $R_0$.
These uncertainties can be written as
\begin{eqnarray}
R_0 & = & \bar{R}_0 (1 \pm \delta_{R_0}) \; ,
\\
R & = & \bar{R} (1 \pm \delta_R) \; .
\end{eqnarray}
The central value $\bar{R}_0$ and the relative error $\delta_{R_0}$ are given in Eq.(\ref{parameter1}), 
$R$ and $\delta_R$ are given by the experimental value from 1993, see Eq. (\ref{REXP}).
Concerning the equality of $R$ and $R_0$:
Eq.(\ref{Master}) shows that the relation between $R$ and $R_0$ reads
\begin{equation}
R = R_0 \left(1+\epsilon \right) \left(1^{+\delta_{+}}_{-\delta_{-}}
\right)\; ,
\end{equation}
with $1+\epsilon = 1/\sqrt{1 - a_{sl}^2}$.
The tiny value of $\epsilon$ can thus be read off from Eq.(\ref{parameter1}), while
$\delta_{+}$ and $\delta_{-}$ can be estimated by
varying $\zeta$ in the range of -1 to +1. One finds sizable values, that clearly cannot
be neglected, because they present by far the dominant uncertainties:
\begin{equation}
\begin{array}{|c||c|c|}
\hline
                    & B_d & B_s
\\
\hline
\hline
\delta_{+} & 0.90  &   0.0027
\\
\hline
\delta_{-} & 0.67 &   0.0028  
\\
\hline
\end{array}
\label{correction}
\; .
\end{equation}
Hence the correctly modified version of Eq.(\ref{zeta}) reads
\begin{equation}
\zeta = \frac{ \left(1+\epsilon \right) 
        \left(1^{+\delta_{+}}_{-\delta_{-}} \right) 
         \sqrt{1-a_{sl}^2} - 1}
        {\alpha - \beta \left(1+\epsilon \right) 
         \left(1^{+\delta_{+}}_{-\delta_{-}} \right)
         \sqrt{1-a_{sl}^2}}
\label{zetacorrect} \; .
\end{equation}
It is evident that Alok and Banerjee have simply set $\epsilon$, $\delta_{+}$ and  $\delta_{-}$
to zero in order to obtain Eq.(\ref{zeta_wrong}).
Taking the finite values of $\epsilon$, $\delta_{+}$ and  $\delta_{-}$ into account 
one gets instead
\begin{equation}
\zeta = 
\frac{0^{+\delta_{+}}_{-\delta_{-}}}
     {\alpha - \beta 
         \left( 1^{+\delta_{+}}_{-\delta_{-}} \right) }
 \; .
\label{zetacorrect2}
\end{equation}
Firstly, we see that the dependence on $a_{sl}$ has disappeared, as it should.
Next, taking the the finite values of $\delta_{+,-}$ from Eq.(\ref{correction})
into account we find using Eq.(\ref{zetacorrect2})
simply that $\zeta$ lies in between -1 and +1, which was what we initially assumed in
order to obtain the values in Eq.(\ref{correction}).
Therefore by rewriting Eq.(\ref{zeta}), in order to avoid using the experimental value of $R$,
one learns nothing new.
\\
To summarise: by neglecting the dominant effect of $\delta_{+,-}$ 
the authors of  \cite{Alok:2013sca} artificially created
a very precise relation, given in  Eq.(\ref{zeta_wrong}), which does not depend 
at all on $R$.  We have shown that this approximation is unjustified and 
leads to a false conclusion.

\section{Conclusion}
\label{sec:conclusion}
In this paper we have investigated decoherence in B-mixing using the most recent values 
for the mixing observables. Within current experimental uncertainties we find no hint 
for any decoherence effect and total decoherence is excluded by about four standard 
deviations 
in the $B_d$-system. The current precision is, however, strongly limited by the
very imprecise
 value of the ratio of like-sign dilepton events to opposite-sign 
dilepton events, $R$.
The most recent experimental number for $R$ stems from 1993. Here any updated measurements, 
using, for example, the large
data set of the B factories, would be very desirable. Moreover, first measurements of $R$ for 
the $B_s$-system (e.g. from the $\Upsilon(5s)$ data set of Belle)  and the $D$-system, e.g. 
from BES would be very interesting.
\\
Finally, we have also shown that the analysis in \cite{Alok:2013sca}, which yields 
very precise bounds on possible decoherence is incorrect, as unjustified assumptions were
made and the dominant uncertainty was simply neglected.

\section*{Acknowledgements}
%\input{acknowledgements}
We would like to thank Guennadi Borissov and Tim Gershon for their
helpful discussion concerning the experimental status and Ashutosh K. Alok
for comments on the manuscript.

  %bibtextest

\end{document}